\documentclass[%
reprint,
superscriptaddress,
nofootinbib,
twocolumn,
notitlepage,
amsmath,amssymb,
aps,
pra
]{revtex4-1}

\usepackage{graphicx}% Include figure files
\usepackage{dcolumn}% Align table columns on decimal point
\usepackage{bm}% bold math
\usepackage{tablefootnote}

\usepackage{amsthm}
\usepackage{amsmath}
\usepackage{amssymb}
\usepackage{graphicx}
\usepackage{url}
\usepackage{float}
\usepackage{bm}
\usepackage{ifthen}
\usepackage[usenames,dvipsnames]{color}
\usepackage{mathrsfs}
\usepackage[colorlinks=true,citecolor=blue,urlcolor=black]{hyperref}
\usepackage{float}
\usepackage{subfigure}
\usepackage{enumitem}
\usepackage{color}
\usepackage{setspace}
\usepackage{braket}
\usepackage{comment}
\usepackage{lipsum}
\newcommand{\be}{\begin{equation}}
\newcommand{\ee}{\end{equation}}

\begin{document}

\title{Machine Learning for Optimal Parameter Prediction in Quantum Key Distribution}

	\author{Wenyuan Wang}
	\affiliation{Centre for Quantum Information and Quantum Control (CQIQC), Dept. of Electrical \& Computer Engineering and Dept. of Physics, University of Toronto, Toronto,  Ontario, M5S 3G4, Canada}

	\author{Hoi-Kwong Lo}
	\affiliation{Centre for Quantum Information and Quantum Control (CQIQC), Dept. of Electrical \& Computer Engineering and Dept. of Physics, University of Toronto, Toronto,  Ontario, M5S 3G4, Canada}
	
\begin{abstract}
	For a practical quantum key distribution (QKD) system, parameter optimization - the choice of intensities and probabilities of sending them - is a crucial step in gaining optimal performance, especially when one realistically considers finite communication time. With the increasing interest in the field to implement QKD over free-space on moving platforms, such as drones, handheld systems, and even satellites, one needs to perform parameter optimization with low latency and with very limited computing power. Moreover, with the advent of the Internet of Things (IoT), a highly attractive direction of QKD could be a quantum network with multiple devices and numerous connections, which provides a huge computational challenge for the controller that optimizes parameters for a large-scale network. Traditionally, such an optimization relies on brute-force search, or local search algorithms, which are computationally intensive, and will be slow on low-power platforms (which increases latency in the system) or infeasible for even moderately large networks. In this work we present a new method that uses a neural network to directly predict the optimal parameters for QKD systems. We test our machine learning algorithm on hardware devices including a Raspberry Pi 3 single-board-computer (similar devices are commonly used on drones) and a mobile phone, both of which have a power consumption of less than 5 watts, and we find a speedup of up to 100-1000 times when compared to standard local search algorithms. The predicted parameters are highly accurate and can preserve over 95-99\% of the optimal secure key rate. Moreover, our approach is highly general and not limited to any specific QKD protocol.

\end{abstract}

\date{\today}
\maketitle

\section{Background}

\subsection{Parameter Optimization in QKD}

Quantum key distribution (QKD)\cite{bb84,e91,QKD_security,QKD_review} provides unconditional security in generating a pair of secure key between two parties, Alice and Bob. To address imperfections in realistic source and detectors, decoy-state QKD \cite{decoystate_LMC,decoystate_Hwang,decoystate_Wang} uses multiple intensities to estimate single-photon contributions, and allows the secure use of Weak Coherent Pulse (WCP) sources, while measurement-device-independent QKD (MDI-QKD) \cite{mdiqkd} addresses susceptibility of detectors to hacking by eliminating detector side channels and allowing Alice and Bob to send signals to an untrusted third party, Charles, who performs the measurement.

In reality, a QKD experiment always has a limited transmission time, therefore the total number of signals is finite. This means that, when estimating the single-photon contributions with decoy-state analysis, one would need to take into consideration the statistical fluctuations of the observables: the Gain and Quantum Bit Error Rate (QBER). This is called the finite-key analysis of QKD. When considering finite-size effects, the choice of intensities and probabilities of sending these intensities is crucial to getting the optimal rate. Therefore, we would need to perform optimizations for the search of parameters. 

Traditionally, the optimization of parameters is implemented as either a brute-force global search for smaller number of parameters, or local search algorithms for larger number of parameters. For instance, in several papers studying MDI-QKD protocols in symmetric \cite{mdiparameter} and asymmetric channels \cite{mdi7intensity}, a local search method called coordinate descent algorithm is used to find the optimal set of intensity and probabilities.

\begin{figure}[t]
	\includegraphics[scale=0.2]{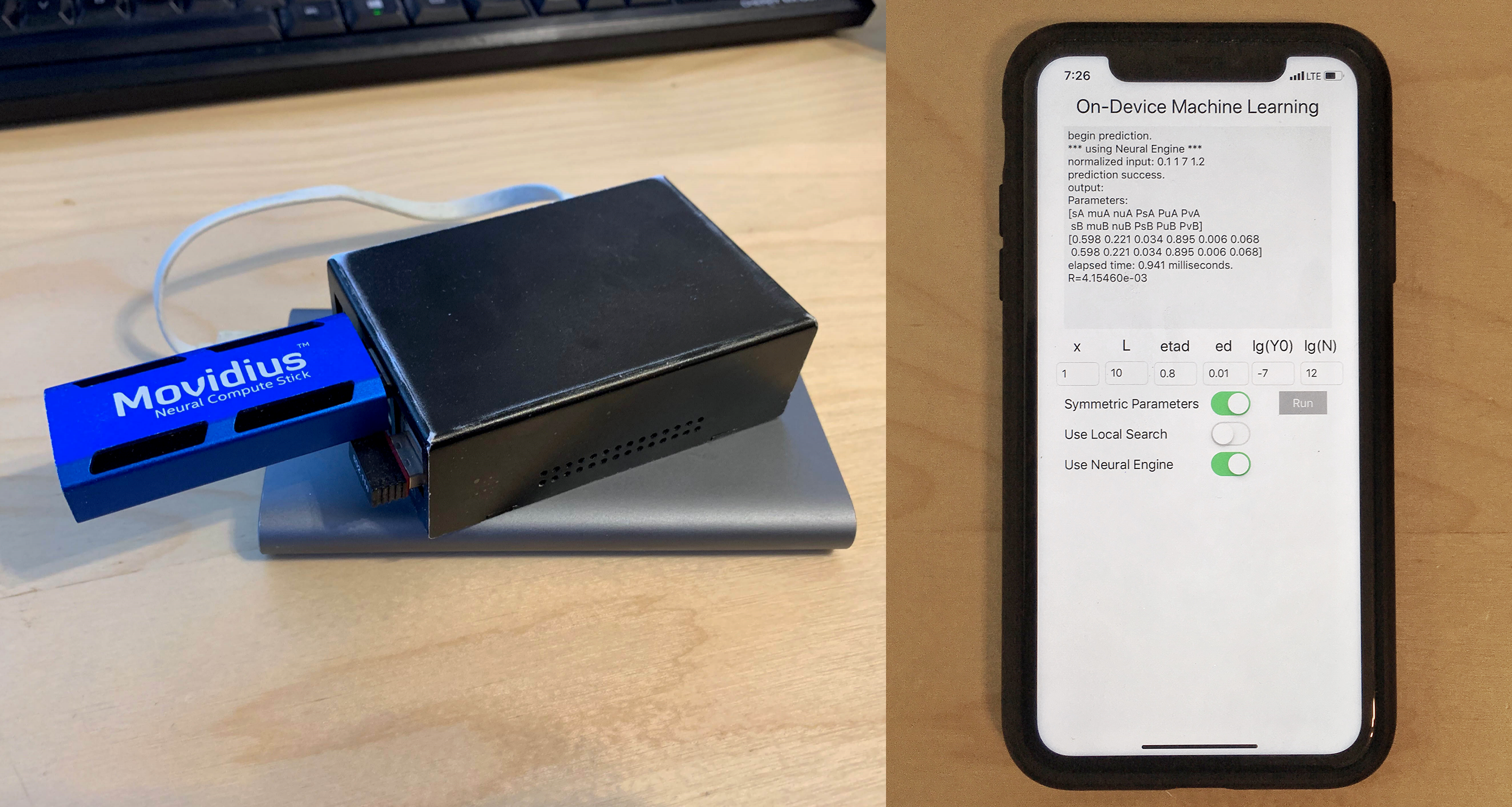}
	\caption{Left: Raspberry Pi 3 single-board computer equipped with Intel Movidius Neural Compute Stick. Right: A smartphone (iPhone XR) running parameter prediction for a QKD protocol with an on-device neural network. In the same app one can also choose to run local search on the device (and compare its running time to that of neural networks).}
	\label{fig:mobile}
\end{figure} 

\begin{table*}[t]
	\begin{minipage}{\textwidth}
		\caption[Caption for LOF]{Time benchmarking between previous local search algorithm and our new algorithm using neural network (NN) for parameter optimization on various devices. Here as an example we consider a protocol for symmetric MDI-QKD \cite{mdifourintensity}. Devices include a Desktop PC with i7-4790k quad-core CPU equipped with a Nvidia Titan Xp GPU, a modern mobile phone Apple iPhone XR with an on-board neural engine, and a low-power single-board computer Raspberry Pi 3 with quad-core CPU, equipped with an Intel Movidius neural compute stick \footnote{The CPU on an iPhone XR has dual big cores + four small cores, but here for simplicity we use a single-threaded program for local search, since OpenMP multithreading library is not supported on Apple devices. OpenMP is supported on the PC and on Raspberry Pi 3, so multithreading is used for local search on these devices.}. As can be seen, neural network generally can provide over 2-3 orders of magnitude higher speed than local search, enabling millisecond-level parameter optimization. Moreover, note that the smartphone and single-board computer provide similar performance with only less than $1/70$ the power consumption, making them ideal for free-space QKD or a quantum internet-of-things. More details on the benchmarking are provided in Section IV.}
		\begin{center}
			\begin{tabular}{ccccc}			
				Device & NN Accelerator & local search  & NN & power consumption\\
				\hline
				Desktop PC & Titan Xp GPU & 0.1s & 0.6-0.8ms & $\sim$350w\\
				iPhone XR & on-board neural engine & 0.2s & 1ms & $<$5w\\
				Raspberry Pi 3 & Intel neural compute stick & 3-5s & 2-3ms & $<$5w\\
			\end{tabular}
		\end{center}
	\end{minipage}
\end{table*}

However, optimization of parameters often require significant computational power. This means that, a QKD system either has to to wait for an optimization off-line (and suffer from delay), or use sub-optimal or even unoptimized parameters in real-time. Moreover, due to the amount of computing resources required, parameter optimization is usually limited to relatively powerful devices such as a desktop PC.

There is increasing interest to implement QKD in free-space on mobile platforms, such as drones, handheld systems, and even satellites. Such devices (e.g. single-board computers and mobile system on chips) are usually limited in computational power. As low-latency is important in such free-space applications, fast and accurate parameter optimization based on a changing environment in real time is a difficult task on such low-power platforms.

Moreover, with the advent of the internet of things (IoT), a highly attractive future direction of QKD is a quantum network that connects multiple devices, each of which could be portable and mobile, and numerous connections are present at the same time. This will present a great computational challenge for the controller of a quantum network with many pairs of users (where real-time optimization might simply be infeasible for even a moderate number of connections).

With the development of machine learning technologies based on neural networks in recent years, and with more and more low-power devices implementing on-board acceleration chips for neural networks, here we present a new method of using neural networks to help predict optimal parameters efficiently on low-power devices. We test our machine learning algorithm in real-life devices such as a single-board computer and a smart phone (see Fig. \ref{fig:mobile}), and find that with our method they can easily perform parameter optimization in milliseconds, within a power consumption of less than 5 watts. We list some time benchmarking results in Table I. Such a method makes it possible to support \textit{real-time} parameter optimization for free-space QKD systems, or large-scale QKD networks with thousands of connections.

\subsection{Neural Network}

\begin{figure}[h]
	\includegraphics[scale=0.33]{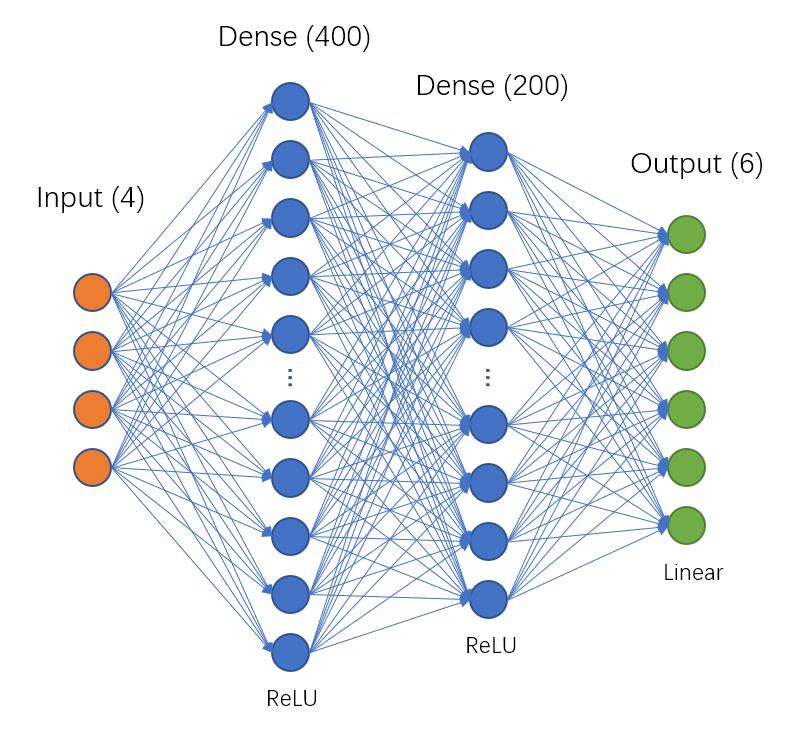}
	\caption{An example of a neural network (in fact, here it is an illustration of the neural network used in our work). It has an input layer and output layer of 4 and 6 neurons, respectively, and has two fully-connected ``hidden" layers with 400 and 200 neurons with rectified linear unit (ReLU) function as activation. The cost function (not shown here) is mean squared error.}
	\label{fig:NN}
\end{figure} 

In this subsection we present a very brief introduction to machine learning with neural networks.

Neural networks are multiple-layered structures built from ``neurons", which simulate the behavior of biological neurons in brains. Each neuron takes a linear combination of inputs $x_i$, with weight $w_i$ and offset $b$, an calculates the activation. For instance:

\begin{equation}
	\sigma(\sum w_i x_i+b)={1\over{1+e^{-(\sum w_i x_i+b)}}}
\end{equation}

\noindent where the example activation function is a commonly used sigmoid function $\sigma(x)={1\over{1+e^{-x}}}$, but it can have other forms, such as a rectified linear unit (ReLU) \cite{RELU} function $max(0,\sum w_i x_i+b)$, a step function, or even a linear function $y=x$.

Each layer of the neural network consists of many neurons, and after accepting input from previous layer and calculating the activation, it outputs the signals to the next layer. Overall, the effect of the neural network is to compute an output $\vec{y}=N(\vec{x})$ from the vector $\vec{x}$. A ``cost function" (e.g. mean square error) is defined on the output layer by comparing the network's calculated output $\{\vec{y}^i\}=\{N(\vec{x_0^i})\}$ on a set of input data $\{\vec{x_0^i}\}$, versus the desired output $\{\vec{y_0^i}\}$. It uses an algorithm called ``backpropagation"\cite{BP} to quickly solve the partial derivatives of the cost function to the internal weights in the network, and adjusts the weights accordingly via an optimizer algorithm such as stochastic gradient descent (SGD) to minimize the cost function and let $\{\vec{y^i}\}$ approach $\{\vec{y_0^i}\}$ as much as possible. Over many iterations, the neural network will be able to learn the behavior of $\{\vec{x_0^i}\}\rightarrow{\vec{y_0^i}}$, so that people can use it to accept a new incoming data $\vec{x}$, and predict the corresponding $\vec{y}$. The \textit{universal approximation theorem} of neural network \cite{Universal} states that it is possible to infinitely approximate any given bounded, continuous function on a given defined domain with a neural network with even just a single hidden layer, which suggests that neural networks are highly flexible and robust structures that can be used in a wide range of scenarios where such mappings between two finite input/output vectors exist.

There is an increasing interest in the field in applying machine learning to improve the performance of quantum communication. For instance, there is recent literature that e.g. apply machine learning to continuous-variable (CV) QKD to improve the noise-filtering \cite{CVQKD1} and the prediction/compensation of intensity evolution of light over time \cite{CVQKD2}, respectively. 

In this work, we apply machine learning to predict the optimal intensity and probability parameters for QKD (based on given experimental parameters, such as channel loss, misalignment, dark count, and data size), and show that with a simple fully-connected neural network with two layers, we can very accurately and efficiently predict parameters that can achieve over 95-99\% the key rate. 

Our work demonstrates the feasibility of deploying neural networks on actual low-power devices, to make them perform fast QKD parameter optimization in real time, with up to 2-3 orders of magnitudes higher speed. This enables potential new applications in free-space or portable QKD devices, such as on a satellite\cite{freespace_satellite1}, drone \cite{freespace_drone}, or handheld \cite{freespace_handheld} QKD system, where power consumption of devices is a crucial factor and computational power is severely limited, and traditional CPU-intensive optimization approaches based on local or global search is infeasible.

Additionally, we point out that with the higher optimization speed, we can also enable applications in a large-scale quantum internet-of-things (IoT) where many small devices can be interconnected (thus generating a large number of connections), and now with neural networks, even low-power devices such as a mobile phone will be able to optimize the parameters for hundreds of users in real-time.

Our paper is organized as follows: In Section II we will describe how we can formulate parameter optimization as a function that can be approximated by a neural network. We then describe the structure of the neural network we use, and how we train it such that it learns to predict optimal parameters. In Section III we test our neural network approach with three example protocols, and show that neural networks can accurately predict parameters, which can be used to obtain near-optimal secure key rate for the protocols. In Section IV we describe two important use cases for our method: enabling real-time parameter optimization on low-power and low-latency portable devices, and paving the road for large-scale quantum networks. We conclude our paper in Section V.

\section{Methods}

In this section we describe the process of training and validating a neural network for parameter optimization. As mentioned in Sec. I, the \textit{universal approximation theorem} implies that the approach is not limited for any specific protocol. Here for simplicity, in this section when describing the methods we will first use a simple symmetric ``4-intensity MDI-QKD protocol" \cite{mdifourintensity} as an example protocol. Later in the next section when presenting the numerical results, we also include other two protocols, the asymmetric ``7-intensity" MDI-QKD protocol\cite{mdi7intensity}, and the finite-size BB84 protocol\cite{finitebb84} to show the method applies to them effectively too. 

\subsection{Optimal Parameters as a Function}

Let us consider the symmetric-channel case for MDI-QKD. Alice and Bob have the same distance to Charles, hence they can choose the same parameters. The variables here will be a set of 6 parameters, $[s, \mu, \nu, P_s, P_\mu, P_\nu]$ for finite-size parameter optimization, where $s, \mu, \nu$ are the signal and decoy intensities, and $P_s, P_\mu, P_\nu$ are the probabilities of sending them. Since only signal intensity $s$ in the Z basis is used for key generation, and $\mu, \nu$ in X basis are used for parameter estimation, $P_s$ is also the basis choice probability. We will unite these 6 parameters into one parameter vector $\vec{p}$.

The calculation of the key rate depends not only on the intensities and the probabilities, but also on the experimental parameters, namely the distance $L$ between Alice and Bob, the detector efficiency $\eta_d$, the dark count probability $Y_0$, the basis misalignment $e_d$, the error-correction efficiency $f_e$, and the number of signals $N$ sent by Alice. We will unite these parameters into one vector $\vec{e}$, which we call the ``experimental parameters".

Therefore, we see that the QKD key rate can be expressed as 

\begin{equation}
Rate=R(\vec{e},\vec{p})
\end{equation}

However, this only calculates the rate for a given fixed set of intensities and experimental parameters. To calculate the optimal rate, we need to calculate

\begin{equation}
R_{max}(\vec{e})=max_{\vec{p} \in P} R(\vec{e},\vec{p})
\end{equation}

\noindent which is the optimal rate. Also, by maximizing R, we end up with a set of optimal parameters $\vec{p_{opt}}$. Note that $\vec{p_{opt}}$ is a function of $\vec{e}$ only, and key objective in QKD optimization is to find the optimal set of $\vec{p_{opt}}$ based on the given $\vec{e}$:

\begin{equation}
\vec{p_{opt}}(\vec{e}) = argmax_{\vec{p} \in P} R(\vec{e},\vec{p})
\end{equation} 

Up so far, the optimal parameters are usually found by performing local or global searches \cite{mdiparameter,mdi7intensity}, which evaluates the function $R(\vec{e},\vec{p})$ many times with different parameters to find the maximum. However, we make the key observation that the functions $R_{max}(\vec{e})$ and $\vec{p_{opt}}(\vec{e})$ are still  \textit{single-valued, deterministic functions} (despite that their mathematical forms are defined by $max$ and $argmax$ and not analytically attainable).\\

As mentioned in Section I, the universal approximation theorem of neural network states that it is possible to infinitely approximate any given bounded, continuous function on a given defined domain with a neural network (with a few or even a single hidden layer). Therefore, this suggests that it might be possible to use a neural network to fully described the behavior of the aforementioned optimal parameter function $\vec{p_{opt}}(\vec{e})$. Once such a neural network is trained, it can be used to directly find the optimal parameter and key rate based on any input $\vec{e}$ by evaluating  $\vec{p_{opt}}(\vec{e})$ and $R(e,\vec{p_{opt}})$ once each (rather than the traditional approach of evaluating the function $R(\vec{e},\vec{p})$ many times) and greatly accelerate the parameter optimization process.

\subsection{Design and Training of Network}

\begin{figure}[h]
	\includegraphics[scale=0.3]{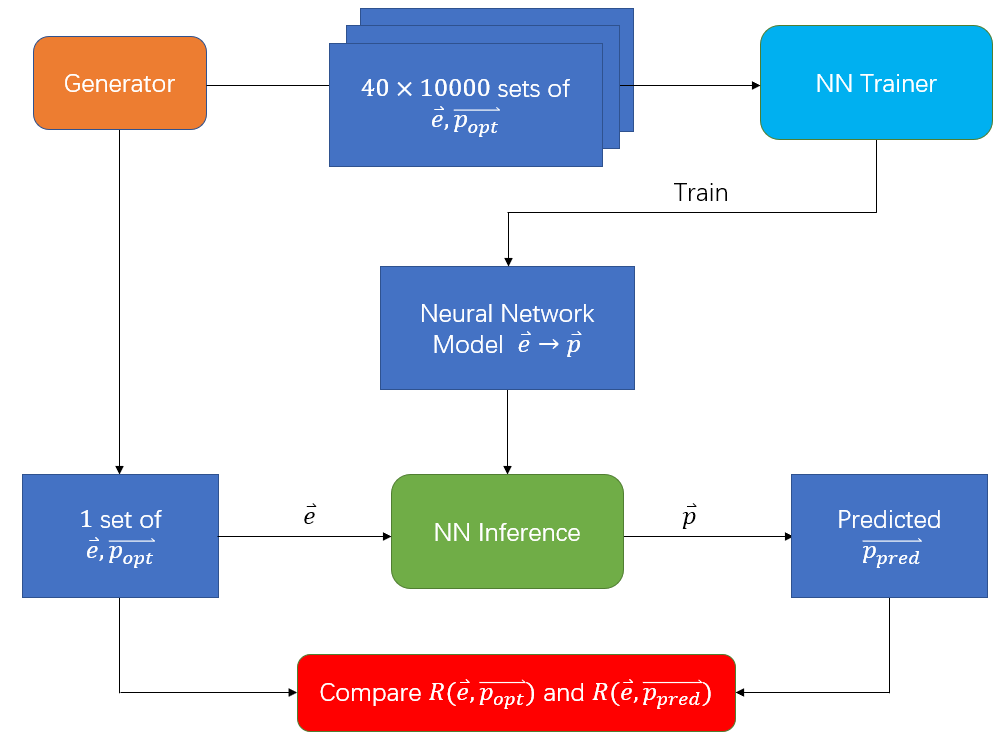}
	\caption{Data flow of training and testing of the neural network (NN). The rounded boxes are programs, and squared boxes represent data. The generator program generates many random sets of experimental parameters $\vec{e}$ and calculates the corresponding optimal parameters $\vec{p_{opt}}$. These data are used to train the neural network. After the training is complete, the network can be used to predict on arbitrary new sets of random experimental data and generate $\vec{p_{pred}}$ (for instance, to plot the results of Fig. 3, a single random set of data is used as input). Finally, another program calculates the key rate based on the actual optimal parameters $\vec{p_{opt}}$ found by local search and the predicted $\vec{p_{pred}}$ respectively, and compare their performances.}
	\label{fig:dataflow}
\end{figure}

\begin{figure*}[t]
	\includegraphics[scale=0.18]{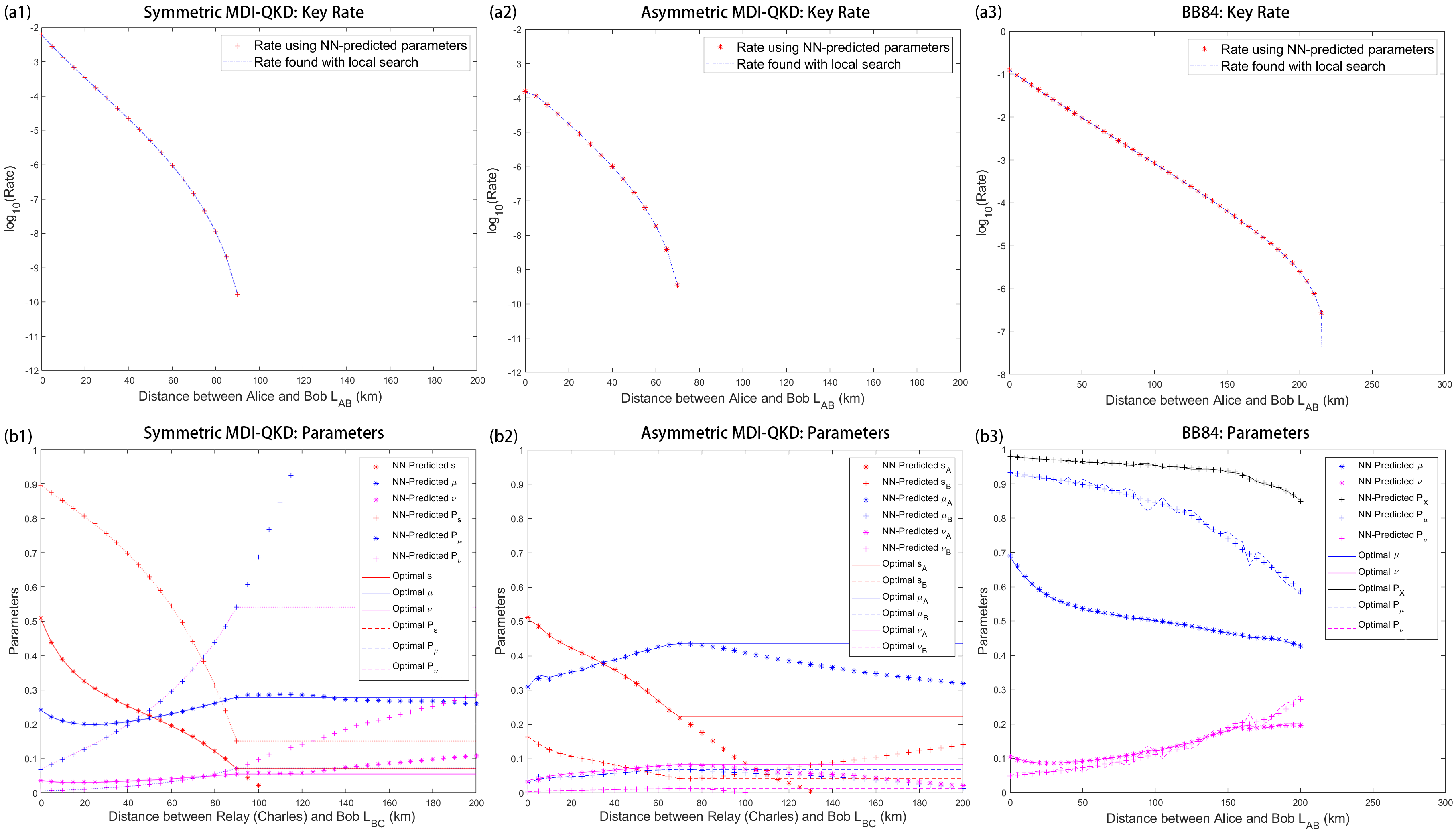}
	\caption{Comparison of neural network predicted parameters vs optimal parameters found by local search for various protocols, using parameters in Table III, at different distances between Alice and Bob. The comparison of neural network (NN) predicted parameters (dots) versus optimal parameters found by local search (lines) are shown in (b1,b2,b3), while the key rate generated with either sets of parameters (dots with NN-predicted parameters, and lines with local search generated parameters) are plotted in (a1,a2,a3). We tested three protocols: (1) symmetric MDI-QKD (4-intensity protocol), (2) asymmetric MDI-QKD (7-intensity protocol), and (3) BB84 protocol. As can be seen, both the NN-predicted parameters and the corresponding key rate are very close to optimal values found with local search. Note that, since the NN does not have any training data on how to choose parameters after key rate is zero, it starts to output arbitrary results after the point where $R=0$, but this does not affect the accuracy of the NN predictions since we are only interested in regions where $R>0$. Also, note that there are some noise present for BB84 protocol. This is because the key rate function shows some level of non-convexity, and we combined local search with a randomized approach (similar to global search) that chooses results from multiple random starting points. Therefore there is some level of noise for the probability parameters (which are insensitive to small perturbations), while the neural network is shown to learn the overall shape of the global maximum of the parameters, and returns a smooth function.}
	\label{fig:rate_param}
\end{figure*} 

Here we proceed to train a neural network to predict the optimal parameters. We first write a program that randomly samples the input data space to pick a random combination of $\vec{e}$ experimental parameters, and use local search algorithm \cite{mdiparameter} to calculate their corresponding optimal rate and parameters. The experimental parameter - optimal parameter data sets (for which we generate 10000 sets of data for 40 points from $L_{BC}=$0-200km, over the course of 6 hours) are then fed into the neural network trainer, to let it learn the characteristics of the function  $\vec{p_{opt}}(\vec{e})$. The neural network structure is shown in Fig.\ref{fig:NN}. With 4 input and 6 output elements, and two hidden layers with 200 and 400 ReLU neurons each. We use a mean squared error cost function. 

For input parameters, since $\eta_d$ is physically no different from the transmittance (e.g. having half the $\eta_d$ is equivalent to having 3dB more loss in the channel), here as an example we fix it to $80\%$ to simplify the network structure (so the input dimension is 4 instead of 5) - when using the network for inference, a different $\eta_d$ can be simply multiplied onto the channel loss while keeping $\eta_d=80\%$. We also normalize parameters by setting 

\begin{equation}
	\begin{aligned}
	e_1&=L_{BC}/100\\
	e_2&=-log_{10}(Y_0)\\
	e_3&=e_d\times 100\\
	e_4&=log_{10}(N)\\
	\end{aligned}
\end{equation}

\noindent to keep them at a similar order of amplitude of 1 (which the neural network is most comfortable with) - what we're doing is a simple scaling of inputs, and this pre-processing doesn't modify the actual data. The output parameters (intensities and probabilities) are within $(0,1)$ to begin with (we don't consider intensities larger than 1 since these values usually provide poor or zero performance) so they don't need pre-processing.

We can also easily modify the setup to accommodate for other protocols by adjusting the number of input and output parameters. For the asymmetric MDI-QKD scenario, one can add an additional input parameter $x=\eta_A/\eta_B$ where $\eta_A,\eta_B$ are the transmittances in Alice's and Bob's channels. We can normalize the mismatch too and make it an additional input variable:

\begin{equation}
\begin{aligned}
e_5&=-log_{10}(x)\\
\end{aligned}
\end{equation}

\noindent In this case the output parameter vector $\vec{p}$ would be $[s_A, \mu_A, \nu_A, P_{s_A}, P_{\mu_A}, P_{\nu_A}, s_B, \mu_B, \nu_B, P_{s_B}, P_{\mu_B}, P_{\nu_B}]$.\\

For finite-size BB84, the input vector is the same as in symmetric MDI-QKD, while the output parameter vector $\vec{p}$ would be $[\mu, \nu, P_\mu, P_\nu, P_X]$ where vacuum+weak decoy states are used (i.e. intensities are $[\mu,\nu,\omega]$)and only one basis - for instance the X basis - is used for encoding. Here $P_X$ is the probability of choosing the X basis.

We trained the network using Adam \cite{ADAM} as the optimizer algorithm for 120 epochs (iterations), which takes roughly 40 minutes on an Nvidia TITAN Xp GPU.

\section{Numerical Results}

\begin{table*}[t]
	\caption{Optimal parameters found by local search vs neural network (NN) predicted parameters for symmetric MDI-QKD using three different random data sets, at the same distance $L_{BC}$ of 20km between Alice and Bob. $Y_0$ is the dark count probability, $e_d$ is the basis misalignment, and $N$ is the number of signals sent by Alice. Here for simplicity, the detector efficiency is fixed at $\eta_d=80\%$ (since it is equivalent to channel loss). Fibre loss per km is assumed to be $\alpha=0.2 dB/km$, the error-correction efficiency is $f_e=1.16$, and finite-size security failure probability is $\epsilon=10^{-7}$. As can be seen, the predicted parameters from our neural network algorithm are very close to the actual optimal parameters, within an $1\%$ error. Moreover, the key rate is even closer, where the rate calculated with predicted parameters can achieve up to 99.99\% the rate found by local search.}
	\begin{center}
		\begin{tabular}{cc|cccc|cccccc}			
			Method & R &$L_{BC}$ &$Y_0$ & $e_d$ & N & $s$ & $\mu$ & $\nu$ & $P_s$ & $P_{\mu}$ & $P_{\nu}$ \\
			\hline\hline
			Local search & $1.3335\times 10^{-3}$& 20km & $1.28 \times 10^{-7}$ & 0.0123 & $1.29 \times 10^{13}$ & 0.501& 0.166&0.0226 & 0.911& 0.00417& 0.0589\\
			NN & $1.3333\times 10^{-3}$ & 20km & $1.28 \times 10^{-7}$ & 0.0123 & $1.29 \times 10^{13}$ & 0.502& 0.167&0.0229 & 0.912& 0.00414& 0.0579\\
			
			Local search & $1.5195\times 10^{-3}$ & 20km & $3.87 \times 10^{-6}$ & 0.0101 & $6.44 \times 10^{12}$ &0.541 &0.179& 0.0256& 0.904 & 0.00480 & 0.0636\\
			NN & $1.5194\times 10^{-3}$ & 20km & $3.87 \times 10^{-6}$ & 0.0101 & $6.44 \times 10^{12}$ &0.542& 0.179 & 0.0257& 0.903&0.0473 & 0.0633\\
			
			Local search & $3.7519\times 10^{-4}$ & 20km & $7.62 \times 10^{-7}$ & 0.0190 & $3.94 \times 10^{11}$ &0.346& 0.212 & 0.0336 & 0.792 & 0.0123 & 0.136\\
			NN & $3.7517\times 10^{-4}$ & 20km & $7.62 \times 10^{-7}$ & 0.0190 & $3.94 \times 10^{11}$ &0.346 & 0.212 & 0.336 & 0.793 & 0.0120 & 0.135\\
			
		\end{tabular}
	\end{center}
\end{table*}

\begin{table}[t]
	\caption[]{Parameters used for simulation of Fig.\ref{fig:rate_param}. $Y_0$ is the dark count probability, $e_d$ is the basis misalignment, and $N$ is the number of signals sent by Alice (and Bob, in MDI-QKD). Here for simplicity, the detector efficiency is fixed at $\eta_d=80\%$. The asymmetry x for MDI-QKD is the ratio of transmittances between Alice's and Bob's channels,$\eta_A/\eta_B$.}
	\begin{center}
		\begin{tabular}{c|ccccc}			
			Parameter Set & x & $e_d$ & $Y_0$  & $N$ & $\eta_d$\\
			\hline
			4-intensity & 1 & 0.014 & $6.2\times 10^{-7}$ & $2.5\times 10^{12}$ & 80\% \\
			%7-intensity & 0.88 & 0.028 & $2.3\times 10^{-7}$ & $5.3\times 10^{13}$ & 80\% \\
			7-intensity & 0.15 & 0.019 & $5.7\times 10^{-6}$ & $3.8\times 10^{12}$ & 80\% \\
			BB84 & - & 0.024 & $1.2\times 10^{-6}$ & $2.8\times 10^{12}$ & 80\% \\
		\end{tabular}
	\end{center}
\end{table}

After training is complete, we use the trained network for 4-intensity protocol to take in three sets of random data, and record the results in Table II. As can be seen, the predicted parameters and the corresponding key rate are very close to the actual optimal values obtained by local search, with the NN-predicted parameters achieving up to 99.99\% the optimal key rate. 

Here we also fix one random set of experimental parameters as seen in Table III, and scan the neural network over $L_{BC}=$0-200km. The results are shown in Fig.\ref{fig:rate_param}(a). As we can see, again the neural network works extremely well at predicting the optimal values for the parameters, and achieves very similar levels of key rate compared to the traditional local search method.

We also use a similar approach to select a random set of input parameters and compare predicted key rate versus optimal key rate for 7-intensity (asymmetric MDI-QKD) protocol, and for finite-size BB84. The results are included in Fig.\ref{fig:rate_param}(b)(c). As can be seen, the accuracy of neural network is very high in these cases too, with up to 95-99\% the key rate for 7-intensity protocol, and up to 99.99\% for finite-size BB84.

\section{Applications and Benchmarking}

In the previous section we have demonstrated that a neural network (NN) can be trained to very accurately simulate the optimal parameter function $\vec{p_{opt}}(\vec{e})$, and be used in effectively predicting the optimal parameters for QKD. The question is, since we already have an efficient coordinate descent (CD) algorithm, what is the potential use for such an NN-prediction program? Here in this section, we will discuss two important use cases for the neural network.\\

\textbf{1. Real-time optimization on low-power devices}. While neural networks take considerable computing power to ``train" (e.g. on a dedicated GPU), using it to predict (commonly called ``inference") is computationally much cheaper, and will be much faster than local search, even if the neural network is run on the same CPU. Moreover, in recent years, with the fast development and wide deployment of neural networks, many manufacturers have opted to develop dedicated chips that accelerate NN-inference on mobile low-power systems. Such chips can further improve inference speed with very little required power, and can also offload the computing tasks from the CPU (which is often reserved for more crucial tasks, such as camera signal processing or motor control on drones, or system operations and background apps on cell phones). 

Therefore, it would be more power-efficient (and much faster) to use an NN-program running on inference chips, rather than using the computationally intensive local search algorithm with CPU on low-power devices. This can be especially important for free-space QKD scenarios such as drone-based, handheld, or satellite-ground QKD, which not only have very limited power budget, but also requires low latency in real-time (for instance, waiting 3-5 seconds for a single-board computer to perform local search each time the channel loss changes - while also using up all CPU resource - would be non-ideal for drones and handheld systems, while a neural network running on a separate accelerator chip at the order of milliseconds would be an ideal choice for such real-time applications.)

As an example, we tested our neural networks on two popular mobile low-power platforms: a single-board computer, and a common mobile phone, as shown in Fig. \ref{fig:mobile}. We implement both CPU-based local search algorithm and neural network prediction on the devices, and list the running time in Table I where we compare neural networks to local search, on the portable devices and on a powerful workstation PC. As shown in Table I, using neural network acceleration chips, we can perform optimization in milliseconds \footnote{Note that, for neural networks it generally takes some time to load the model into the device when first used (about 0.2-0.3s on Titan Xp GPU and neural engine on the iPhone, and 3s on Raspberry Pi with the neural compute stick), but this only needs to be done once at boot time, and can be considered part of the startup time of the device - once the network is running, the predictions can be performed on many sets of data taking only milliseconds for each operation.} (which is 2-3 orders of magnitude faster than CPU local search), in a power footprint less than$1/70$ that of a workstation PC.

In Table I we used 4-intensity protocol as one example, although note that for other protocols, e.g. 7-intensity protocol, the advantage of NN still holds, since the neural network prediction time is little affected by the input/output size (for instance, in Fig. \ref{fig:NN}, there are $400\times 200$ connections between the two middle hidden layers, and only $4\times 400$ and $6\times 200$ connections involving output or input neurons. This means that the numbers of input/output nodes have little impact on the overall complexity of the network), while local search time increases almost linearly with the number of output parameters. For instance, running 7-intensity protocol, which has 12 output parameters, takes about 0.4s using local search on an iPhone XR - which is double the time for 4-intensity protocol, which has 6 output parameters - but with a NN it still takes about 1ms (making the advantage of using NN even greater in this case).

Additionally, note that even without neural network acceleration chips, many devices can still run the neural network on CPU (at the expense of some CPU resource), and this option is still much faster than local search (for instance, running neural network on iPhone XR with CPU takes between $1.3-2.0$ms, which is not much slower than the dedicated neural accelerator chip).\\

%\protect\footnotemark[1]
\textbf{2. Quantum networks}. In addition to free-space QKD applications which require low-power, low-latency devices, the neural network can also be very useful in a network setting, such as a quantum internet-of-things (IoT) where numerous small devices might be interconnected in a network as users or relays. For an untrusted relay network, MDI-QKD protocol is desirable. However, the number of pairs of connections between users will increase quadratically with the number of users, which might quickly overload the compute resources of the relay/users.

With the neural network, any low-power device such as a single-board computer or a mobile phone can easily serve as a relay that connects to numerous users and optimizes 10000 pairs of connections (100 users) in under 10 seconds. This is a task previously unimaginable even for a desktop PC, which can barely optimize parameters for 10 users in the same amount of time using local search. Therefore, our new method can greatly lower the bar for compute power of devices and reduce the latency of the systems when building a quantum Internet of Things.

\section{Conclusion and Discussions}

In this work we have presented a simple way to train a neural network that accurately and efficiently predicts the optimal parameters for a given QKD protocol, based on the characterization of devices and channels. We show that the approach is general and not limited to any specific form of protocol, and demonstrate its effectiveness for three examples: symmetric/asymmetric MDI-QKD, and finite-size BB84.

We show that an important use of such an approach is to enable efficient parameter optimization on low-power devices. We can achieve 2-3 orders of magnitude faster optimization speed compared to local search, with a fraction of the power consumption. It can be run on either the increasingly popular neural network acceleration chips, or on common CPUs that have relatively weak performance. This can be highly useful not only for free-space QKD applications that require low latency and have low power budget, but also for a quantum internet-of-things (IoT) where even a small portable device connected to numerous users can easily optimize all parameters in real-time.

Here we have demonstrated that the technique of machine learning can indeed be used to optimize the performance of a QKD protocol. The effectiveness of this simple demonstration suggests that it may be possible to apply similar methods to other optimization tasks, which are common in the designing and control of practical QKD systems, such as determining the optimal threshold for post-selection in free-space QKD, tuning the polarization controller motors for misalignment control, etc.. We hope that our work can further inspire future works in investigating how machine learning could help us in building better performing, more robust QKD systems.\\

\textit{Note added:} After our posting of a first draft of this work on the preprint server \cite{arxiv}, another work on a similar subject was subsequently posted on the preprint server \cite{arxiv_JOSAB_AI} and later published at \cite{JOSAB_AI}. While both our work and the other work \cite{arxiv_JOSAB_AI,JOSAB_AI} have similar approaches in parameter optimization with neural networks, and observe the huge speedup neural network has over CPU local search, a few important differences remain. Firstly, we show that the neural network method is a general approach not limited to any specific protocols (and show its versatile applications with three examples), while Ref. \cite{arxiv_JOSAB_AI,JOSAB_AI} is limited to discussing asymmetric MDI-QKD only. Secondly, we point out that a key use case of this approach would be performing parameter optimization on low-power devices with neural networks. This was only briefly mentioned in passing in Ref. \cite{arxiv_JOSAB_AI,JOSAB_AI}. In contrast, we perform testing and benchmarking on real hardware devices. Our work not only will allow more types of smaller portable devices to join a network setting, but also is important in free-space QKD applications where power consumption is crucial.

\section{Acknowledgments}

This work was supported by the Natural Sciences and Engineering Research Council of Canada (NSERC), U.S. Office of Naval Research (ONR). We sincerely thank Nvidia for generously providing a Titan Xp GPU through the GPU Grant Program.


\begin{thebibliography}{}
		\bibitem{bb84} C Bennett, G Brassard, `` Quantum cryptography: Public key distribution and coin tossing." International Conference on Computer System and Signal Processing, IEEE (1984).
		\bibitem{e91} AK Ekert, ``Quantum cryptography based on Bell’s theorem." Physical review letters 67.6:661 (1991).
		\bibitem{QKD_security} P Shor, J Preskill, "Simple proof of security of the BB84 quantum key distribution protocol." Physical review letters 85.2 (2000): 441.
		\bibitem{QKD_review} N Gisin, G Ribordy, W Tittel, H Zbinden, ``Quantum cryptography." Reviews of modern physics 74.1:145 (2002).
		\bibitem{decoystate_Hwang} WY Hwang, ``Quantum key distribution with high loss: toward global secure communication." Physical Review Letters 91.5 (2003): 057901.
		\bibitem{decoystate_LMC} HK Lo, XF Ma, and K Chen, ``Decoy state quantum key distribution." Physical review letters 94.23 (2005): 230504.
		\bibitem{decoystate_Wang} XB Wang, ``Beating the photon-number-splitting attack in practical quantum cryptography." Physical review letters 94.23 (2005): 230503.
		\bibitem{mdiqkd} HK Lo, M Curty, and B Qi, ''Measurement-device-independent quantum key distribution." Physical review letters 108.13 (2012): 130503.
		
		\bibitem{mdiparameter} F Xu, H Xu, and HK Lo, ``Protocol choice and parameter optimization in decoy-state measurement-device-independent quantum key distribution." Physical Review A 89.5 (2014): 052333.
		\bibitem{mdi7intensity} W Wang, F Xu, and HK Lo. ``Enabling a scalable high-rate measurement-device-independent quantum key distribution network." arXiv preprint arXiv:1807.03466 (2018).
		\bibitem{RELU} V Nair and GE Hinton. ``Rectified linear units improve restricted boltzmann machines." Proceedings of the 27th international conference on machine learning (ICML-10). 2010.		
		\bibitem{BP} R Hecht-Nielsen, ``Theory of the backpropagation neural network." Neural networks for perception. 1992. 65-93.
		\bibitem{Universal} K Hornik, MStinchcombe, and H White. ``Multilayer feedforward networks are universal approximators." Neural networks 2.5 (1989): 359-366.	
		\bibitem{CVQKD1} W Lu, C Huang, K Hou, L Shi, H Zhao, Z Li, J Qiu, ``Recurrent neural network approach to quantum signal: coherent state restoration for continuous-variable quantum key distribution." Quantum Information Processing 17.5 (2018): 109.
		\bibitem{CVQKD2} W Liu, P Huang, J Peng, J Fan, G Zeng, ``Integrating machine learning to achieve an automatic parameter prediction for practical continuous-variable quantum key distribution." Physical Review A 97.2 (2018): 022316.
		
		\bibitem{freespace_satellite1} S-K Liao et al. ``Satellite-to-ground quantum key distribution." Nature 549.7670 (2017): 43-47.
		\bibitem{freespace_drone} AD Hill, J Chapman, K Herndon, C Chopp, DJ Gauthier, P Kwiat, ``Drone-based Quantum Key Distribution", QCRYPT 2017 (2017).
		\bibitem{freespace_handheld} G Mélen, P Freiwang, J Luhn, T Vogl, M Rau, C Sonnleitner, W Rosenfeld, and H Weinfurter, ``Handheld Quantum Key Distribution." Quantum Information and Measurement. Optical Society of America (2017).
		
		\bibitem{mdifourintensity} YH Zhou, ZW Yu, and XB Wang, ``Making the decoy-state measurement-device-independent quantum key distribution practically useful." Physical Review A 93.4 (2016): 042324.		
		\bibitem{finitebb84} CCW Lim, M Curty, N Walenta, F Xu and H Zbinden, ``Concise security bounds for practical decoy-state quantum key distribution." Physical Review A 89.2 (2014): 022307.
		\bibitem{ADAM} DP Kingma and LJ Ba, ``Adam: A method for stochastic optimization." arXiv preprint arXiv:1412.6980 (2014).
		
		\bibitem{arxiv} W Wang, HK Lo, ``Machine Learning for Optimal Parameter Prediction in Quantum Key Distribution." arXiv preprint arXiv:1812.07724 (2018).
		\bibitem{arxiv_JOSAB_AI} FY Lu, et al. ``Parameter optimization and real-time calibration of measurement-device-independent quantum key distribution network based on back propagation artificial neural network." arXiv preprint arXiv:1812.08388 (2018).
		\bibitem{JOSAB_AI} FY Lu, et al. ``Parameter optimization and real-time calibration of a measurement-device-independent quantum key distribution network based on a back propagation artificial neural network." JOSA B 36.3: B92-B98 (2019).
		
		
		
		
\end{thebibliography}
\end{document}